\documentclass[twocolumn,secnumarabic,amssymb, nobibnotes, aps, prl, superscriptaddress]{revtex4-2}

\usepackage{graphicx}
\usepackage{dcolumn}
\usepackage{bm}
\usepackage{amsmath}
\usepackage{enumerate}
\usepackage{setspace}
\usepackage{dsfont}
\usepackage{subfigure}
\usepackage{multirow}
\usepackage{indentfirst} 
\usepackage {mathrsfs}
\usepackage{color}

\usepackage{lineno}

\usepackage{threeparttable}  
\usepackage[pdfstartview=FitH,
CJKbookmarks=true,
bookmarksnumbered=true,
bookmarksopen=true,
colorlinks, 
pdfborder=001,
linkcolor=blue,
anchorcolor=blue,
citecolor=blue
]{hyperref}

\begin{document}
	

\title{Continuous-mode quantum key distribution with digital signal processing}%
\author{Ziyang Chen}%
\affiliation{State Key Laboratory of Advanced Optical Communication Systems and Networks, School of Electronics, and Center for Quantum Information Technology, Peking University, Beijing 100871, China}

\author{Xiangyu Wang}%
\affiliation{State Key Laboratory of Information Photonics and Optical Communications, Beijing University of Posts and Telecommunications, Beijing 100876, China}

\author{Song Yu}%
\affiliation{State Key Laboratory of Information Photonics and Optical Communications, Beijing University of Posts and Telecommunications, Beijing 100876, China}

\author{Zhengyu Li}%
\email[E-mail: ]{lizhengyu2@huawei.com}
\affiliation{Huawei Technologies Co., Ltd., Shenzhen 518129, China}

\author{Hong Guo}%
\email[E-mail: ]{hongguo@pku.edu.cn}
\affiliation{State Key Laboratory of Advanced Optical Communication Systems and Networks, School of Electronics, and Center for Quantum Information Technology, Peking University, Beijing 100871, China}

\date{\today}%

\begin{abstract}
	Continuous-variable quantum key distribution (CVQKD) offers the specific advantage of sharing keys remotely by the use of standard telecom components, thereby promoting cost-effective and high-performance metropolitan applications. Nevertheless, the introduction of high-rate spectrum broadening has pushed CVQKD from a single-mode to a continuous-mode region, resulting in the adoption of modern digital signal processing (DSP) technologies to recover quadrature information from continuous-mode quantum states. However, the security proof of DSP involving multi-point processing is a missing step. Here, we propose a generalized method of analyzing continuous-mode state processing by linear DSP via temporal-modes theory. The construction of temporal modes is key in reducing the security proof to single-mode scenarios. 
The proposed practicality oriented security analysis method paves the way for building classical compatible digital CVQKD.

\end{abstract}

\maketitle

\section{Introduction}

Quantum key distribution (QKD)~\cite{Rev.Mod.Phys.74.145.2002,Adv.Opt.Photon.12.1012.2020,Rev.Mod.Phys.92.025002.2020} promises an information-theoretically secure symmetric key distribution for distant partners. The past three decades have witnessed rapid development of QKD technologies and the growth of QKD network deployment globally, which have been employed in various security applications~\cite{New.J.Phys.11.075001.2009,CLEO2011,Opt.Express.26.24260.2018,IEEE.Communications.Surveys.Tutorials.21.881.2019,npj.Quantum.Inf.5.101.2019,Sci.Adv.6.eaba0959.2020,Nat.Photon.15.850.2021,Nature.589.214.2021}. Within the QKD family, continuous-variable (CV) QKD  benefits from the use of off-the-shelf commercial telecom components~\cite{Entropy.17.6072.2015,Adv.Quantum.Technol.1.1800011.2018} and provides a cost-effective alternative in metropolitan networks. Twenty years since the pioneering GG02 protocol~\cite{Phys.Rev.Lett.88.057902.2002} was proposed, the theory~\cite{Phys.Rev.Lett.97.190503.2006,Phys.Rev.Lett.109.100502.2012,Phys.Rev.Lett.114.070501.2015,Phys.Rev.Lett.118.200501.2017,Phys.Rev.Research.3.013279.2021,Phys.Rev.Research.3.043014.2021} and experimentation~\cite{Nat.Photon.7.378.2013,Sci.Rep.6.19201.2016,Phys.Rev.Lett.125.010502.2020} of CVQKD have made remarkable progress.

Moreover, the tremendous breakthroughs of local oscillator (LO) schemes since 2015~\cite{Phys.Rev.X.5.041009.2015,Phys.Rev.X.5.041010.2015} have pushed CVQKD into a new stage, in which techniques from modern digital coherent communication have been brought in~\cite{IEEE.Photon.Technol.Lett.30.650.2018,Commun.Phys.2.9.2019,J.Lightwave.Technol.38.2214.2020,Commun.Phys.5.162.2022}. We call this stage \emph{digital} CVQKD. Specifically, digital signal processing (DSP) significantly improves the signal-to-noise ratio (SNR) by compensating for channel drifting and device impairments, which greatly simplifies physical systems. This paves the way for an ultra-high secret key rate with tens of GHz of bandwidth in the CVQKD system. However, this phenomenon also complicates the security analysis.

Two barriers exist in the security analysis of a digital CVQKD system. One is the discrete modulation format, and the other is DSP. The former results from the destruction of estimating the covariance matrix directly from the measurement results 
and was recently solved with the semidefinite programming~\cite{Phys.Rev.X.9.021059.2019,Phys.Rev.X.9.041064.2019,Quantum.5.540.2021} or other novel methods~\cite{arXiv:1805.04249,Nat.Commun.12.252.2021}. 
The other barrier is the difficulty of constructing an appropriate measurement operator to describe the output of DSP.



Specifically, a single-point quadrature measurement of each state in one ensemble is sufficient for reliable tomography of single-mode states. However, for the tomography of continuous-mode states, the extraction of quadrature information involves multi-point sampling and processing. Therefore, a time-domain description of system's behavior should be introduced, which is beyond the traditional single-mode description.



\begin{figure*}[t!]
	\centering
	\includegraphics[width=1 \linewidth]{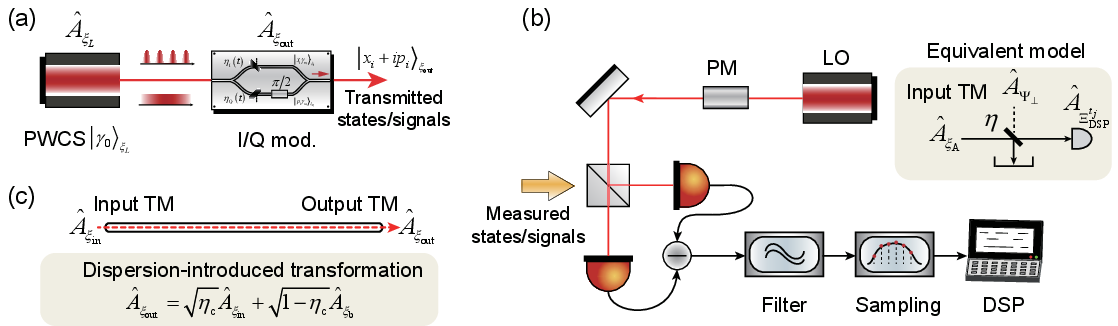}
	\caption{Models of digital CVQKD. (a) The transmitter prepares a photon-wavepacket coherent state (PWCS) with an arbitrary form of envelope ${{\xi _L}\left( t \right)}$; via the I/Q modulation, the quadratures of the PWCS are modulated by Gaussian distributed random numbers. The carrier can be either a continuous wave or pulsed coherent state. (b) Measured states are fed to a practical receiver, interfering with the local oscillator (LO) at a balanced beam splitter (BS) and detected by a band-limited homodyne detector (HD) (modeled by an ideal HD and a filter), followed by the sampling and DSP devices. The mismatch between the measured state's temporal mode (TM) and the receiver's TM is equivalent to a BS with a transmitivity of $\eta$. (c) In the experiment, channel transmission causes a transformation of TM, which, if ignored, will cause a degradation in system performance. PM: phase modulator.}
	\label{Protocol}
\end{figure*}

 Here, to narrow the gap between practical systems and theoretical models, we develop a generalized security proof framework for continuous-mode systems processed by linear DSP algorithms. The key step is the temporal-mode (TM) construction using DSP results, which is suitable for the analysis of high-speed, multi-point sampling systems. Specifically, in continuous-mode formalism, time-domain field operators can be introduced by Fourier transformation, based on which the generalized receiver can be well modeled. By properly calibrating the shot-noise unit (SNU), we model the linear processing of sampled data by recombination of time-domain field operators, which defines a specific TM field operator~\cite{Phys.Rev.A.42.4102.1990,Rev.Mod.Phys.92.40.2020,Phys.Scr.95.064002.2020}. Consequently, the security of DSP is reduced to the security of a specific single TM measurement. Then, the rest of the analysis is compatible with traditional methods.

Moreover, the results show that the mismatch between the measured state's TM and the receiver's TM leads to inefficiency in detection. The mission of the DSP algorithm is to merge this mismatch, thus improving the detection efficiency, which coincides with improving SNR in its classical correspondence.

Our work provides a feasible way of analyzing the security and performance of a continuous-mode system processed by linear DSP algorithms, so it could provide important guidance for the DSP design of digital CVQKD. Linear DSP toolboxes are expected to be directly employed in CVQKD, reinforcing the importance of our work.

\section{Results}

\textbf{Temporal modes of continuous-mode states}.

We start by introducing the basics of continuous-mode formalism of quantum optics and then describing the state preparation phase.
Recall that in traditional CVQKD analysis, the coherent state is represented by the creation and annihilation operators in terms of a single-mode field, given by $\hat a_i^\dag ,{\hat a_i}$. By contrast, in a practical system, high-speed modulation inevitably introduces a nonuniform temporal waveform, so continuous-mode formalism of field operators~\cite{Phys.Rev.Lett.63.1586.1989,Phys.Rev.A.42.4102.1990,Phys.Scr.95.064002.2020} should be introduced, which is widely used in studying continuous-mode quantum optics. By transforming the annihilation and creation operators from their discrete-mode counterparts, the continuous-mode field operators are defined as ${\hat a_i} \to {\left( {\Delta \omega } \right)^{\frac{1}{2}}}\hat a\left( \omega  \right)$ and ${\hat a_i^{\dag}} \to {\left( {\Delta \omega } \right)^{\frac{1}{2}}}\hat a^{\dag}\left( \omega  \right)$, where ${\Delta \omega }$ denotes the mode spacing, which satisfies the commutation relation $\left[ {\hat a\left( \omega  \right),{{\hat a}^\dag }\left( {\omega '} \right)} \right] = \delta \left( {\omega  - \omega '} \right)$. 
In the time domain, it is useful to define the Fourier transforms of $ \hat a\left( \omega  \right)$, namely $\hat a\left( t \right)$, given by $\hat a\left( t \right) = \frac{1}{{\sqrt {2\pi } }}\int {d\omega } \hat a\left( \omega  \right)\exp \left( { - i\omega t} \right)$. The creation operator $\hat a^\dag \left( t \right)$ follows a similar definition.

Based on this, the photon-wavepacket creation operator $\hat A_{{\xi _i}}^\dag$~\cite{Phys.Rev.A.42.4102.1990,QuantumTheoryofLight} can be defined as  
\begin{equation}\label{WavePacketOperator}
\hat A_{{\xi _i}}^\dag = \int {dt } {\xi _i}\left( t  \right){\hat a^\dag }\left( t  \right),
\end{equation}
in which the wavepacket ${\xi _i}\left( t \right)$ usually reads $\xi_i^0\left(t\right)e^{-i\omega t}$, as an envelope $\xi_i^0\left(t\right)$ with a carrier~$e^{-i\omega t}$. 
It is also known as the TM field operator~\cite{Rev.Mod.Phys.92.40.2020,Phys.Scr.95.064002.2020} if ${\xi _i}\left( t \right)$ meets the orthonormalization that $\int {dt} \xi _i^*\left( t \right){\xi _j}\left( t \right) = {\delta _{ij}}$, for different $i,j$. The TM operators also obey the commutation relation, which reads 
\begin{equation}\label{CommutationRelationWP}
\left[ {{{\hat A}_{{\xi _i}}},\hat A_{{\xi _j}}^\dag } \right] = {\delta _{ij}}.
\end{equation}
It is then important to define the photon-wavepacket coherent state ${\left| {{\gamma _i}} \right\rangle _{{\xi _i}}}$ on $\xi_i$-TM, as~\cite{Phys.Rev.A.42.4102.1990}
\begin{equation}\label{WPCS}
{\left| {{\gamma _i}} \right\rangle _{{\xi _i}}} = \hat D_{\xi_i} \left(\gamma_i\right) \left| 0 \right\rangle=  \exp \left( {{\gamma _i}\hat A_{{\xi _i}}^\dag  - \gamma _i^*{{\hat A}_{{\xi _i}}}} \right)\left| 0 \right\rangle ,
\end{equation}
where ${\gamma _i}$ denotes the displacement parameter, and ${\left| {{\gamma _i}} \right|^2}$ represents the average photon number. The photon-wavepacket coherent state obeys the eigenvalue equation ${{\hat A}_{{\xi _i}}}{\left| {{\gamma _i}} \right\rangle _{{\xi _i}}} = {\gamma _i}{\left| {{\gamma _i}} \right\rangle _{{\xi _i}}}$.
Under this notation, the quadrature operator with the phase angle $\theta$ can be defined as
\begin{equation}\label{Quadrature_WPCS}
\hat X_{{\xi _i}}^\theta  = \hat A_{{\xi _i}}^\dag \exp \left( {i\theta } \right) + {{\hat A}_{{\xi _i}}}\exp \left( { - i\theta } \right).
\end{equation}


In a digital CVQKD system, coherent states are generated by widely used in-phase/quadrature (I/Q) modulation. As shown in Fig.~\ref{Protocol}(a), assuming that ${\left| {{\gamma _0}} \right\rangle _{{\xi _{L}}}}$ is the photon-wavepacket coherent state fed to the I/Q modulator, ${\left| {{\gamma _m}} \right\rangle _{{\xi _{L}}}}$ are then the state of the I or Q arm after the balanced beam splitter (BS), where ${\gamma _m} = {{{\gamma _0}} \mathord{\left/{\vphantom {{{\gamma _i}} {\sqrt 2 }}} \right.\kern-\nulldelimiterspace} {\sqrt 2 }}$. Then, each arm performs the intensity modulation with a certain waveform, which is modeled by a time-dependent BS with transmitivity ${\eta \left( t \right)}$ related to modulation~\cite{AGuidetoExperimentsinQuantumOptics,J.Mod.Opt.34.881.1987}. Assuming that the data encoded on the I and Q components in the $i$-th period are $\left\{ {{x_i}} \right\}$ and $\left\{ {{p_i}} \right\}$ and that their normalized waveform envelopes are ${\xi _{\rm{I}}}\left( t \right)$ and ${\xi _{\rm{Q}}}\left( t \right)$, we can rewrite $\sqrt {{\eta _{\rm{I}}}\left( t \right)}\xi_L^0\left(t\right)  = {x_i}{\xi _{\rm{I}}}\left( t \right)$ on the I path and $\sqrt {{\eta _{\rm{Q}}}\left( t \right)}\xi_L^0\left(t\right)  = {p_i}{\xi _{\rm{Q}}}\left( t \right)$ on the Q path. Then, the I and Q path's output states are transformed into ${\left| {{x_i}{\gamma _m}} \right\rangle _{{\xi _{\rm{I}}}}}$ and ${\left| {{p_i}{\gamma _m}} \right\rangle _{{\xi _{\rm{Q}}}}}$.
Finally, after passing through another balanced BS and a proper attenuator, the output photon-wavepacket coherent state is $\left| {{x_i} + i{p_i}}  \right\rangle _{{\xi _{{\rm{out}}}}}$ if ${\xi _{\rm{I}}}\left( t \right) = {\xi _{\rm{Q}}}\left( t \right) = {\xi _{{\rm{out}}}}\left( t \right)$ with properly calibrated modulation. 
This means that in the entanglement-based scheme, the output coherent states can be seen as two-mode squeezed states, with $\xi_{\rm{out}}$-TM being measured on one mode with a heterodyne measurement.   
As for ${\xi _{\rm{I}}}\left( t \right) \ne {\xi _{\rm{Q}}}\left( t \right)$, the output state can be decomposed into two orthogonal TMs with different but correlated displacement parameters, which we leave for further investigations.

\textbf{Measurement, sampling, and data processing}.

On the receiver side, the input state is first measured by a practical homodyne detector with limited bandwidth and then sampled by an analog-to-digital converter (ADC). After this, the sampled data go through a series of DSP algorithms, and the final data output from DSP are assumed to represent the quadrature measurement result, which can be used to construct the covariance matrix and then calculate the secret key rate. We only consider linear DSP algorithms here because the transmitted quantum light is extremely weak, so no obvious optical nonlinear effects occur. Thus, linear compensation algorithms are sufficient to recover the signal. 
When we use the above continuous-mode formalism notation, mapping the outputs of linear DSP to quadrature measurements is surprisingly natural; the crucial step is normalization with properly calibrated SNU. To paint a clear picture of this, we first ignore the imperfections of the homodyne detector and finite-resolution issue of ADC, and we discuss the trusted detection model considering the detector’s efficiency and noise in the Methods section.

The receiver can be modeled as an ideal homodyne detector, followed by a filter, as shown in Fig.~\ref{Protocol}(b). Assuming the filter has an impulse response function (IRF), namely, $g\left( t \right)$, the photocurrent flux operator of a homodyne detector is given by~\cite{Phys.Rev.A.42.4102.1990,QuantumTheoryofLight}
\begin{equation}\label{HD_Bandlimited}
	\hat f\left( t \right) = \left[ {{{\hat a}^\dag }\left( t \right){{\hat a}_{{\rm{LO}}}}\left( t \right) + \hat a_{{\rm{LO}}}^\dag \left( t \right)\hat a\left( t \right)} \right]*g\left( t \right),
\end{equation}
where $*$ denotes the convolution. 
The photon wavepacket of the local oscillator (LO) is given by ${\alpha _{{\rm{LO}}}}\left( t \right) = \mu _{{\rm{LO}}}^{{1 \mathord{\left/{\vphantom {1 2}} \right.\kern-\nulldelimiterspace} 2}}{\xi _{{\rm{LO}}}}\left( t \right)\exp \left( {-i{\omega _{{\rm{LO}}}}t + i\theta } \right)$, where ${\mu _{\rm{LO}}}$ denotes the average number of photons contained in an envelope ${\xi _{{\rm{LO}}}}\left( t \right)$ for a pulsed LO, or a time period as defined. 
Because LO is considered a classical field with enough photons, the fluctuation in the measurement output mainly comes from the signal quadrature measurement part. Therefore, the photocurrent flux after taking the average over LO is more useful, given by ${{\hat f}_{{\rm{LO}}}}\left( t \right) = \langle {\alpha _{{\rm{LO}}}}\left( t \right)|\hat f\left( t \right)\left| {{\alpha _{{\rm{LO}}}}\left( t \right)} \right\rangle  $.

Considering ADC as the integral sampling process with integral time $\Delta t_s $, the sampled data at time $t_0$ are 
\begin{equation}\label{samplingResults}
{{\hat D}_{{t_0}}} = \frac{1}{{\Delta {t_{\rm{s}}}}}\int_{{t_0}}^{{t_0} + \Delta {t_{\rm{s}}}} {dt} {{\hat f}_{{\rm{LO}}}}\left( t \right),
\end{equation}
in which the electronics amplification is ignored.

Multiple sampling points may exist within one period~$T_{\rm{r}}$. Generally, types of linear data processing exist: i) directly choose one sampled data of each period as final data of this period, for instance, the sampling point near the peak of the envelope of measured state; ii) using the sampled data within one period to define the final data of this period, for instance, calculating the weighted averaging of all sampled data within the same period; and iii) generally, a DSP algorithm may use the sampled data from multiple periods, for instance, the root raised cosine (RRC) filter \cite{RRC_paper} introduces a convolution over multiple periods. 


For a DSP algorithm involving $N$ sampled data, the output data at the time corresponding to the $t_j$ sampling time could be
\begin{equation}\label{fdsp}
	{{\hat D}_{t_j}^N} = {f_{{\rm{dsp}}}}\left( {\hat D}_{t_{j-k+1} },... {\hat D}_{t_{j-k+N} } \right)= \sum\limits_{i = 1}^N {f_{{\rm{dsp}}}^i {{\hat D}_{{t_{j-k+i}}}} }
\end{equation} 
with linear expansion, where $f_{{\rm{dsp}}}^i$ and $k$ are real numbers determined by DSP algorithms. After simplification, Eq.~(\ref{fdsp}) is given by
\begin{align}	
{{\hat D}_{t_j}^N} \!=\! \frac{\mu _{{\rm{LO}}}^{{1 \mathord{\left/
{\vphantom {1 2}} \right.
\kern-\nulldelimiterspace} 2}}} {{\Delta {t_{\rm{s}}}}}\int  G_{{\rm{dsp}}}^{t_j}\left( \tau \right)\hat X^{\alpha_{\rm{LO}}}\left(\tau\right) {d\tau},
	\label{SAMPLINGResultsRaw}
\end{align}
where
\begin{equation}\label{DSP_DetectionFunction}
G_{{\rm{dsp}}}^{t_j}\left( \tau \right) = \sum\limits_{i = 1}^N {f_{{\rm{dsp}}}^i} \int_{{t_{j-k+i}}}^{{t_{j-k+i}} + \Delta {t_{\rm{s}}}}  g\left( {t - \tau} \right){dt}
\end{equation}
is related to the detector's IRF, sampling points, and the DSP algorithm. In addition, $X^{\alpha_{\rm{LO}}}\left(\tau\right)$ is the intermediate quadrature operator related to the LO's features, which is given by
\begin{equation}\label{quadrature_operator_general}
	\begin{aligned}
{\hat X^{{\alpha _{{\rm{LO}}}}}}\left( \tau  \right) &= \xi_{\rm{LO}}\left(\tau\right)e^{  {-i\left( {{\omega _{{\rm{LO}}}}\tau - \theta } \right)} }{{\hat a}^\dag }\left( \tau \right)  + \rm{h.c.} \\
	\end{aligned}
\end{equation}

\textbf{SNU calibration and normalization}.

To normalize the output data from DSP, one key step is to define and calibrate the SNU, which is the most distinguishable phase from classical optical communication. Considering $\hat D ^N _ {t_j}$ as the final data for the period in which $t_j$ lies, we can easily verify that for the vacuum input, the mean is $\langle 0|{{\hat D}_{t_j}^N}| 0 \rangle=0$ , and the variance $\sigma _{{\rm{SNU}}}^2 = \langle 0|{{\hat D}^N_{t_j}}{{\hat D}^N_{t_j}}\left| 0 \right\rangle $ is
\begin{equation}\label{SNUVariance}
{\sigma^2 _{{\rm{SNU}}}} = \frac{\mu _{{\rm{LO}}}}{{\Delta {t_{\rm{s}}}^2}}\int  \left|{\xi _{{\rm{LO}}}}\left( \tau \right)\right|^2 \left[G_{{\rm{dsp}}}^{t_j}\left( \tau \right)\right]^2{d\tau}.
\end{equation}
For normalization, the sampled data $\hat D_{t_j}^{{\rm{SNU}}}$ are divided by $\sigma _{{\rm{SNU}}}=\sqrt{\sigma^2 _{{\rm{SNU}}}}$, which gives 
\begin{equation}\label{SNU_calibratedResults_middle}
\hat D_{t_j}^{{\rm{SNU}}}\!  =\! e^{i\theta}\int {d\tau}  
\frac{{G_{{\rm{dsp}}}^{t_j}\left( \tau \right)} \xi_{\rm{LO}}\left(\tau\right)e^{-i\omega_{\rm{LO}}\tau}}{{{\sigma _{{\rm{cal}}}}}} \hat a ^{\dag} \left(\tau\right) \! + \! \rm{h.c.},
\end{equation}
where ${\sigma _{{\rm{cal}}}} = \sqrt {\int {d\tau } {{\left| {{\xi _{{\rm{LO}}}}\left( \tau  \right)} \right|}^2}{{\left[ {G_{{\rm{dsp}}}^{{t_j}}\left( \tau  \right)} \right]}^2}}$ is the rescaled factor when calibrating output data by SNU. It can be verified that the coefficient function of $\hat a \left(\tau \right)$ is a normalized photon-wavepacket function, which is 
\begin{equation}\label{IF_TM}
\Xi _{\rm{DSP}}^{t_j}\left( \tau \right) = \frac{1}{{{\sigma _{{\rm{cal}}}}}} {{\xi _{{\rm{LO}}}}\left( \tau \right)G_{{\rm{dsp}}}^{t_j}\left( \tau \right)}\exp \left( {-i{\omega _{{\rm{LO}}}}\tau} \right),
\end{equation}
with the normalization condition $\int {d\tau} {\left| {\Xi _{\rm{DSP}}^{t_j}\left( \tau \right)} \right|^2} = 1$. This introduces $\Xi _{\rm{DSP}}^{t_j}$-TM, which is jointly defined by the LO, filter, sampling, and DSP algorithms. Then, we can further define its creation operator as
\begin{equation}\label{TemporalMode_field_creation}
\hat A_{\Xi _{\rm{DSP}}^{t_j}}^\dag  = \int {d \tau} \Xi _{\rm{DSP}}^{t_j}\left(\tau\right){{\hat a}^\dag }\left( \tau \right).
\end{equation}
Consequently, a simplified form of Eq.~(\ref{SNU_calibratedResults_middle}) in terms of the $\Xi _{\rm{DSP}}^{t_j}$-TM operators is rewritten as 
\begin{equation}\label{SNU_TM}
	\hat D_{{t_j}}^{{\rm{SNU}}} \!=\! \hat A_{\Xi _{{\rm{DSP}}}^{{t_j}}}^\dag \exp \left( {i\theta } \right) \! +\! {\hat A_{\Xi _{{\rm{DSP}}}^{{t_j}}}} \exp \left( { - i\theta } \right) \!=\! \hat X_{\Xi _{{\rm{DSP}}}^{{t_j}}}^\theta,
\end{equation}
which shares the same form as Eq.~(\ref{Quadrature_WPCS}). 

Therefore, the final data (output from DSP and being normalized) can be treated as a quadrature measurement of $\Xi _{\rm{DSP}}^{t_j}$-TM. As long as the data represent a quadrature measurement result, they can be used to construct the covariance matrix and are thus compatible with traditional security analysis methods.    
The abovementioned analysis also highlights one key point of SNU calibration, which is that the sampled data of vacuum input should be processed by the same DSP as the usual signal input case before the data are used to calculate the variance of shot noise. 

Another important issue is that for a DSP involving sampled data exceeding one period, the possible crosstalk should be avoided. In this case, the TMs related to different periods should be orthogonal, that is, $ \int \Xi _{\rm{DSP}}^{t_i} \Xi _{\rm{DSP}}^{t_j*} =0$, where $t_i,t_j$ belong to two different periods.
This also coincides with classical DSP's purpose, as crosstalk lowers the SNR. For instance, the RRC pulse shaping and filtering methods are commonly used to improve spectrum efficiency. Moreover, the RRC filter is designed with no intersymbol interference for different optimal sampling points, which is associated with the TMs related to different optimal sampling points being orthogonal.     

This completes our security analysis framework for linear DSP, which can be summarized in two key points. One is properly calibrating SNU, which naturally leads the final data to be a quadrature measurement result with respect to the TM defined jointly defined by LO, detector, sampling, and DSP. The second is to avoid the complex measurement model introduced by intersymbol crosstalk, where TMs corresponding to different periods should be orthogonal. If these two conditions apply, the final data can be directly used to construct the covariance matrix, and then we can calculate the secret key rate through the current security analysis method.

Below, two examples are given to analyze the performance after considering the continuous-mode scenario, including the mode-matching issue and transmission-dispersion issue.


\begin{figure*}[t!]
	\centering
	\includegraphics[width=0.7 \linewidth]{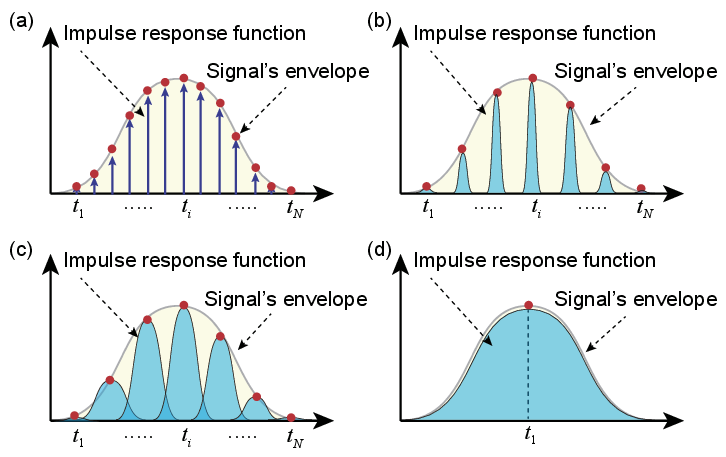}
	\caption{Weighted averaging of sampled data. Assume the LO is a continuous-wave (CW) laser. (a) The bandwidth of the detector is large enough, and the impulse response function (IRF) is approximate to the $\delta $-function. The final data are recovered by weighted averaging of all of the sampled data within one period, and the weight follows the shape of the measured state's envelope. (b), (c) The bandwidth of the detector is limited, resulting in a widened IRF, which convolutes the signal around each sampling point. (d) If the detector's IRF is similar to the measured state's envelope, single-point sampling can represent the final data.}
	\label{fig_Detection}
\end{figure*}

\textbf{Projection on the measurement temporal mode}.

Besides security, our analysis also provides insights into imperfect detection efficiency, related to the mismatch between the measured state's TM and the receiver's TM. The whole measurement process under the TM representation is equivalent to the projection of the measured state's TM to the receiver's TM, and vice versa. 

Now, consider a case where the measured state is a coherent state on $\xi_{\rm{A}}$-TM, which is different from the receiver's $\Xi _{\rm{DSP}}^{t_j}$-TM. Using the Gram--Schmidt orthogonalization, we can define a third TM from $\Xi _{\rm{DSP}}^{t_j}$-TM and orthogonal to $\xi_{\rm{A}}$-TM, denoted as ${{\Psi _ \bot }}$-TM, which leads to the following decomposition of the creation operator:
\begin{equation}\label{TMdecomposed}
\hat A_{\Xi _{{\rm{DSP}}}^{{t_j}}}^\dag  = \sqrt \eta  \hat A_{{\xi _{\rm{A}}}}^\dag  + \sqrt {1 - \eta } \hat A_{{\Psi _ \bot }}^\dag ,
\end{equation}
where $\eta  = {\left[ {\int {dt} \Xi _{{\rm{DSP}}}^{{t_j},*}\left( t \right){\xi _{\rm{A}}}\left( t \right)} \right]^2} \le 1$ denotes the mode-matching coefficient. 
With further examination of the first-order and second-order moments, the abovementioned decomposition can be modeled by an extra BS at the receiver side, with transmitivity $\eta$, quantifying the matching degree between $\xi_{\rm{A}}$-TM and $\Xi _{\rm{DSP}}^{t_j}$-TM. A detailed derivation can be found in the Methods section. Here, $\eta < 1 $ means an extra loss induced by the mode mismatch, which decreases the performance of the system. This degradation is rather covert, different from the physical components introduced by loss, that is, fiber coupling loss and the non-unit quantum efficiency of photodiodes. The closer $\eta$ is to 1, the better the performance of a DSP algorithm.  

Here, we take the weighted averaging scenario to further illustrate the mode-matching issue in the time domain. In this case, the DSP function is given by $f_{{\rm{dsp}}}^i = {w_i}$, where ${w_i}$ is the weight of the $i$-th sampling point within one period. The measurement results can be described by Fig.~\ref{fig_Detection}. A sampled data point measures not only the signal at one time point but also convoluted nearby signals around the sampling point. Therefore, an intuitive understanding of mode matching is that the sum of all sampled data covers a certain signal area. In Fig.~\ref{fig_Detection}(a), the bandwidth of the detector is large enough that the IRF is approximate to the $\delta$-function, and then ultra-dense sampling is needed.  
By contrast, if the bandwidth gradually decreases, the IRF becomes wider, and fewer sampling points are required to achieve a similar mode-matching degree, as shown in Fig.~\ref{fig_Detection}(b), (c). If the detector's IRF is similar to the measured state's envelope, a single sampling point is enough, as in Fig.~\ref{fig_Detection}(d).

\begin{figure*}
	\centering
	\includegraphics[width=0.7 \linewidth]{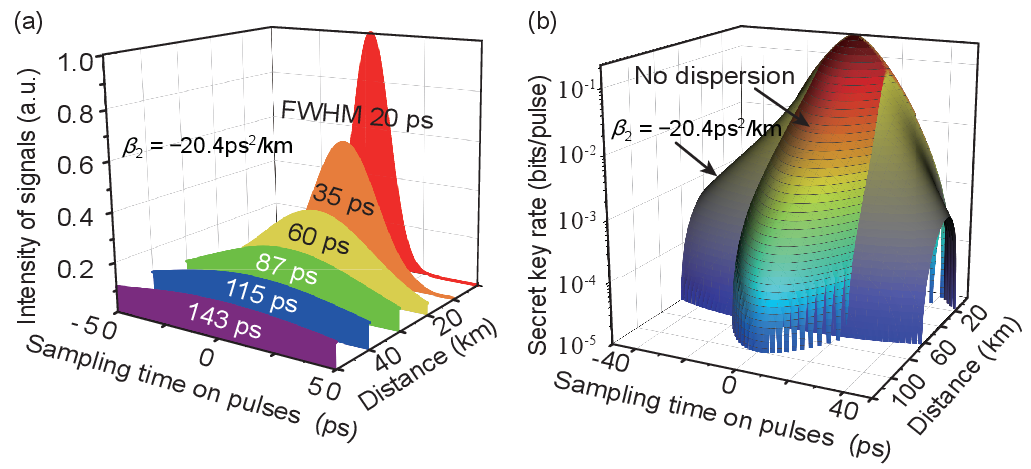}
	\caption{Simulation results of the channel dispersion effect. (a) Evolution of the envelope of a Gaussian pulse with a full width at half-maximum (FWHM) of 20 ps in optical fibers, ignoring the fiber loss, where ${\beta _2} =  - 20.4{\rm{p}}{{\rm{s}}^2}/{\rm{km}}$ is considered, which is a typical value of standard G.652 telecom fiber at 1,550 nm. (b) Secret key rates for narrow-pulse (Gaussian pulse) propagation considering fiber-dispersion-induced TM mismatch, where ${\beta _2} =  0{\rm{p}}{{\rm{s}}^2}/{\rm{km}}$ and ${\beta _2} =  - 20.4{\rm{p}}{{\rm{s}}^2}/{\rm{km}}$ are considered, which are typical values of no dispersion channel and the standard G.652 telecom fiber at 1,550 nm. In the simulation, we use the single-sampling-point scheme, and the sampling time accuracy is considered. The sampling time $t = 0{\rm{ps}}$ is assumed as the peak of Gaussian pulse. It is shown that the increasing mode mismatch reduces the secret key rate while also relaxing the accuracy requirement of the sampling time. }
	\label{Simulation_Results}
\end{figure*}

We also note that, the discussion of system's side information is an interesting research topic. We discuss it in two scenarios: 1) Alice performs a good calibration of
the modulation variance of the transmitted signal; 2) Alice performs a poor calibration of the
modulation variance, for example, neglecting optical modes that may exist on Alice’s side.

For the first scenario, all of the optical modes in the spectrum are taken into account, including the part outside the bandwidth of Bob's detector. Because the information of the whole transmitter is included in the variance of Alice's TM, any energy that Bob does not detect (either the energy beyond the detector's bandwidth or the energy loss caused by the mode mismatch) contributes to the channel loss estimated from covariance matrix, which will be considered as caused by the eavesdropper. Therefore, this is not a side channel, but rather a performance degradation.

For the second scenario, Alice fails to perform a proper calibration, which will open a source-flaw-related side channel. For example, part of the signal's energy of the sideband is not included when calibrating the modulation variance of Alice. This issue has been extensively studied in the single-mode case~\cite{Npj.Quantum.Inf.8.136.2022,Phys.Rev.A.93.032309.2016,Phys.Rev.A.96.062309.2017,Phys.Rev.A.98.062319.2018,Quantum.Sci.Technol.6.045001.2021}. Note that this is not caused by the continuous-mode model proposed by our work, but by the improper calibration method of the modulation variance at the transmitter. To avoid this issue, three alternative methods are proposed: 1) Add a power meter at Alice's side to calibrate Alice's overall energy and determine the modulation variance of the whole TM; 2) avoid information leakage through single-sideband modulation~\cite{Npj.Quantum.Inf.8.136.2022}; 3) take the leakage information into account in the overall security analysis~\cite{Phys.Rev.A.93.032309.2016,Phys.Rev.A.96.062309.2017,Phys.Rev.A.98.062319.2018,Quantum.Sci.Technol.6.045001.2021}.

Therefore, in the practical security analysis of a CVQKD system, especially when the homodyne detector is used to calibrate Alice's modulation variance (such as using Bob's device to perform back-to-back test for engineering convenience), we should deal with this phenomenon more carefully. We will investigate this issue in detail in the future.

\textbf{Continuous-mode transmission.} 

The measured state's TM is determined by the modulation at the transmitter side, which is controlled, and also the channel transmission, which is not controlled. Therefore, the DSP needs to be designed or adjusted according to the channel condition. We take channel dispersion as an example to show how it influences the measurement, which is significant in a high-speed system. 
Considering a coherent state with a narrow Gaussian wavepacket ${\xi _{{\rm{in}}}}\left( t \right)$, after passing through the channel, its output envelope is transformed with a transfer function $h\left( {t,z} \right)$, given by ${\xi _{{\rm{out}}}}\left( t \right) = {\xi _{{\rm{in}}}}\left( t \right)*h\left( {t,z} \right)$,
where $h\left( {t,z} \right)$ is the channel transfer function in the time domain, and its Fourier transform is given by $\mathscr{F}\left[ {h\left( {t,z} \right)} \right]  = \exp \left( { - {k_i}z + i{k_1}\Omega z + i\frac{{{k_2}}}{2}{\Omega ^2}z} \right)$,
in which the second-order Taylor expansion of the real part of the wave vector is considered, $z$ is the transmission distance, $\Omega $ is the Fourier frequency, and ${k_1}$ and ${k_2}$ denote the inverse group velocity and second-order dispersion coefficient, respectively. Only for the ultrashort period, the influence of third-order nonlinear dispersion needs to be considered. It can be seen that the output state's TM varies with increasing distance, as in Fig.~\ref{Simulation_Results}(a). Therefore, if the DSP does not consider this TM varying, there will be an increasing extra loss as the transmission distance increases, which will decrease the maximal transmission distance, as  simulated in Fig.~\ref{Simulation_Results}(b). This is where self-adaptive algorithms apply.


\section{Discussion}

In this study, we have developed a generalized practical system  model with  continuous-mode formalism of quantum optics, based on which the IQ modulation at the transmitter side and band-limited homodyne detection with the sampling process at the receiver side can be well described. Then, with proper calibration of SNU, the output data of a linear DSP can be modeled by the quadrature measurement result with respect to a specific TM, jointly defined by the LO, filter, sampling, and DSP algorithms. This immediately results in good compatibility with traditional security analysis methods, which completes the security proof of linear DSP algorithms. Linear DSP toolboxes are expected to be directly employed in CVQKD, which highlights the importance of our work.

In addition to the security, our work also provides a method to analyze the performance of a DSP algorithm through a factor quantifying the matching degree between the measured state's TM and the receiver's TM. Moreover, interesting concepts like the DSP-induced fast fading-channel effect can be further analyzed to explore the practical limitations of a CVQKD system. By the guidance of our work, secure and better-performing DSP algorithms can be designed, which will exploit the significant potential of digital CVQKD to achieve ultra-high secret-key-generation speed and cost-effective implementations. 


\section{METHODS}

\textbf{Trusted detection model considering DSP.}

\begin{figure}[t]
	\centering
	\includegraphics[width=0.9 \linewidth]{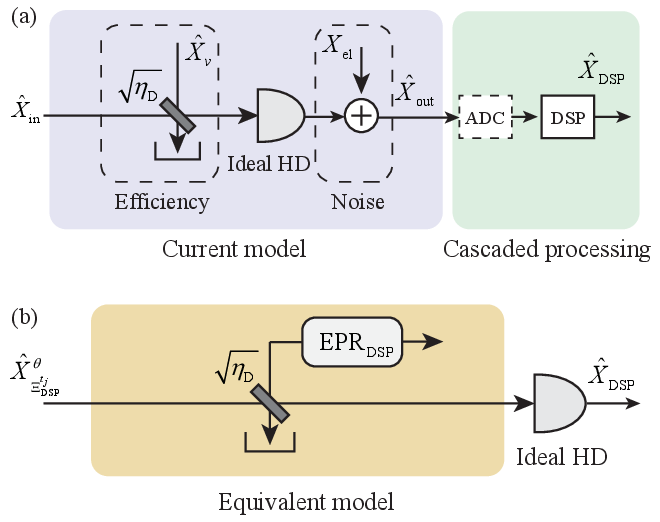}
	\caption{The proposed detection model considering non-ideal efficiency and electronic noise. (a) The PM version of our detection model. (b) The equivalent EB version of our detection model.}
	\label{fig2}
\end{figure}

Our model mainly deals with the DSP part, which is cascaded after the trusted physical detection process. The input of the DSP modular is actually the output of the practical trusted detector, namely, ${{\hat X}_{{\rm{out}}}}$, as shown in Fig.~\ref{fig2} (a). From this point of view, our analysis is actually a complement to the existing trusted model after considering the time-related information and its processing.

Since we restrict the discussed DSP algorithms to linear algorithms, the processing of the output signal is equivalent to the independent processing of the incident signal and the electronic noise. Then the output of DSP shares the same form as the output of the current trusted model, given by
\begin{align}
	&{{\hat X}_{{\rm{DSP}}}}   = {f_{{\rm{dsp}}}}\left( {{{\hat X}_{{\rm{out}}}}} \right)  \notag  \\
	&= \sqrt {{\eta _{\rm{D}}}} {f_{{\rm{dsp}}}}\left( {{{\hat X}_{{\rm{in}}}}} \right) + \sqrt {1 - {\eta _{\rm{D}}}} {f_{{\rm{dsp}}}}\left( {{{\hat X}_v}} \right) + {f_{{\rm{dsp}}}}\left( {{X_{{\rm{el}}}}} \right)  \notag \\
	& \mathop  = \limits^{{\rm{PM}}} \sqrt {{\eta _{\rm{D}}}} \hat X_{\Xi _{{\rm{DSP}}}^{{t_j}}}^\theta  + \sqrt {1 - {\eta _{\rm{D}}}} {{\hat X}_{\Xi _{{\rm{DSP}},v}^{{t_j}}}} + X_{{\rm{el}}}^{{\rm{DSP}}} \label{PM_result} \\
	&\mathop  = \limits^{{\rm{EB}}} \sqrt {{\eta _{\rm{D}}}} \hat X_{\Xi _{{\rm{DSP}}}^{{t_j}}}^\theta  + \sqrt {1 - {\eta _{\rm{D}}}} \hat X_{{\rm{EPR}}}^{{\rm{DSP}}},  \label{EB_result}
\end{align}
where ${f_{{\rm{dsp}}}}\left(  \cdot  \right)$ denotes the linear DSP function, and $\hat X_{\Xi _{{\rm{DSP}}}^{{t_j}}}^\theta $ is the incident TM defined by a specific DSP algorithm. The result of the prepare-and-measure (PM) scheme (Fig.~\ref{fig2} (a)) is given in Eq.~(\ref{PM_result}), where ${{\hat X}_{\Xi _{{\rm{DSP}},v}^{{t_j}}}}$ refers to the TM of a vacuum input, and the equivalent electronic noise $X_{{\rm{el}}}^{{\rm{DSP}}}$ is the broadband electronic noise filtered by the same DSP algorithm. The result of the equivalent entanglement-based (EB) scheme (Fig.~\ref{fig2} (b)) is given in Eq.~(\ref{EB_result}), where $\hat X_{{\rm{EPR}}}^{{\rm{DSP}}}$ is the equivalent trusted mode introduced by our model.

Therefore, the trusted model of a practical detector with DSP can be simplified as Fig.~\ref{fig2} (b), as long as we re-calibrate the variance of the equivalent electronic noise $v_{{\rm{el}}}^{{\rm{DSP}}}$.

In experiments, the practical calibrating steps are actually very similar to the current method, given by the following two steps:

1) Turn off the quantum input signal, turn off the LO, and directly sample the output of the detector, which corresponds to the measurement data of the electronic noise;

2) Process the collected data by the same DSP function as measured signal and then use the processing result to calculate the variance of the electronic noise.

After the calibrated electronic noise variance is obtained, the variance of the trusted EPR mode introduced in the EB scheme can be obtained, given by $V_{{\rm{EPR}}}^{{\rm{DSP}}} = 1 + {{v_{{\rm{el}}}^{{\rm{DSP}}}} \mathord{\left/{\vphantom {{v_{{\rm{el}}}^{{\rm{DSP}}}} {\left( {1 - {\eta _{{\rm{D}}}}} \right)}}} \right. \kern-\nulldelimiterspace} {\left( {1 - {\eta _{{\rm{D}}}}} \right)}}$, referred to as Bob's input.

\textbf{Derivation of the mode-matching coefficient.}

Assume that the measured state at Bob's input is an unknown wavepacket coherent state ${\left| \gamma  \right\rangle _{{\xi _{\rm{A}}}}}$ related to the wavepacket ${\xi _{\rm{A}}}\left( t \right)$. To obtain the equivalent performance of mode matching, we first decompose the receiver's basis function $\Xi _{{\rm{DSP}}}^{{t_j}}\left( t  \right)$  into the input basis ${\xi _{\rm{A}}}\left( t \right)$ and its orthogonal basis ${{\Psi _ \bot }\left( t \right)}$. Then we show the mode-matching coefficient.

Using the Gram--Schmidt process, we can map the receiver's operator from the basis $\Xi _{{\rm{DSP}}}^{{t_j}}\left( t  \right)$  to the set of bases $\left\{ {{\xi _{\rm{A}}}\left( t \right),{\Psi _ \bot }\left( t \right)} \right\}$. In this transformation, the basis $\Xi _{{\rm{DSP}}}^{{t_j}}\left( t  \right)$ represents the measurement mode-matched basis function. The Gram--Schmidt process is given as follows:
\begin{itemize}
	\item Step 1: The overlapping of two bases (also called the mode-matching coefficient) is defined as
	\begin{equation}\label{OverlappingBasis}
		\sqrt \eta   = \int {dt} \Xi _{{\rm{DSP}}}^{{t_j}, * }\left( t \right){\xi _{\rm{A}}}\left( t \right).
	\end{equation}
	\item Step 2: The mode-matched basis is written as ${\zeta _1}\left( t \right) = {\xi _{\rm{A}}}\left( t \right)$ directly.
	\item Step 3: The second orthonormal basis is calculated by
	\begin{align}
		{\zeta _2}\left( t \right) & = \Xi _{{\rm{DSP}}}^{{t_j}}\left( t \right) - \frac{{\int {dt} \Xi _{{\rm{DSP}}}^{{t_j}, * }\left( t \right){\zeta _1}\left( t \right)}}{{\int {dt} \zeta _1^*\left( t \right){\zeta _1}\left( t \right)}}{\zeta _1}\left( t \right)\notag \\
		&=\Xi _{{\rm{DSP}}}^{{t_j}}\left( t \right)  - \sqrt \eta  {\xi _{\rm{A}}}\left( t \right). 
	\end{align}
\end{itemize}
It is easy to verify that
\begin{align}
	&\int {dt} \zeta _2^*\left( t \right){\zeta _2}\left( t \right)   \notag \\
	&= \int {dt} \left[ {\Xi _{{\rm{DSP}}}^{{t_j}, * }\left( t \right) - \sqrt \eta  \xi _{\rm{A}}^*\left( t \right)} \right]\left[ {\Xi _{{\rm{DSP}}}^{{t_j}}\left( t \right) - \sqrt \eta  {\xi _{\rm{A}}}\left( t \right)} \right]\notag \\
	& = 1 + \eta  - 2\sqrt \eta  \int {dt{\xi _{\rm{A}}}\left( t \right)\Xi _{{\rm{DSP}}}^{{t_j}}\left( t \right)}   \notag \\
	&= 1 - \eta.
\end{align}
After normalizing the basis function, we can obtain
\begin{align}
	{{\Psi _ \bot }\left( t \right)} &= \frac{{{\zeta _2}\left( t \right)}}{{\sqrt {\int {dt} \zeta _2^ * \left( t \right){\zeta _2}\left( t \right)} }} \notag\\
	& = \frac{1}{{\sqrt {1 - \eta } }}\left( {\Xi _{{\rm{DSP}}}^{{t_j}}\left( t \right) - \sqrt \eta  {\xi _{\rm{A}}}\left( t \right)} \right).
\end{align}
The receiver's basis function is then decomposed as
\begin{equation}\label{}
	\Xi _{{\rm{DSP}}}^{{t_j}}\left( t \right) = \sqrt \eta  {\xi _{\rm{A}}}\left( t \right) + \sqrt {1 - \eta } {\Psi _ \bot }\left( t \right),
\end{equation}
and based on this, we can define two temporal modes (TMs) given by
\begin{align}
	\hat A_{{\xi _{\rm{A}}}}^\dag  &= \int {dt} {\xi _{\rm{A}}}\left( t \right){{\hat a}^\dag }\left( t \right),  \\
	\hat A_{{\Psi _ \bot }}^\dag  &= \int {dt} {\Psi _ \bot }\left( t \right){{\hat a}^\dag }\left( t \right).
\end{align}

Now, the measurement results can be rewritten as
\begin{align}
	\hat D_{{t_j}}^{{\rm{SNU}}} = \hat X_{\Xi _{{\rm{DSP}}}^{{t_j}}}^\theta = \sqrt \eta  {{\hat X}_{{\xi _{\rm{A}}}}} + \sqrt {1 - \eta } {{\hat X}_{{\Psi _ \bot }}}, \label{MeasurementResult_ATT}
\end{align}
where ${{\hat X}_\xi } = {{\hat A}_\xi } + \hat A_\xi ^\dag$ denotes the quadrature operator of $\xi$-TM.

\textbf{Moments of Measured Data.}

Now, let us investigate the first-order and second-order moments of the final data. The mean value (first-order moment) is given by
\begin{align}
	d_{\rm{out}} &\! = \!\langle \gamma |\hat D_{{t_j}}^{{\rm{SNU}}}{\left| \gamma  \right\rangle _{{\xi _{\rm{A}}}}} \!=\! \langle \gamma |\left( {\sqrt \eta  {{\hat X}_{{\xi _{\rm{A}}}}} + \sqrt {1 - \eta } {{\hat X}_{{\Psi _ \bot }}}} \right){\left| \gamma  \right\rangle _{{\xi _{\rm{A}}}}}   \notag \\
	&= \sqrt \eta  \langle \gamma |{{\hat X}_{{\xi _{\rm{A}}}}}{\left| \gamma  \right\rangle _{{\xi _{\rm{A}}}}} + \sqrt {1 - \eta } \langle \gamma |{{\hat X}_{{\Psi _ \bot }}}{\left| \gamma  \right\rangle _{{\xi _{\rm{A}}}}}  \notag \\
	&= \sqrt \eta  {d_{\rm{in}}},	 \label{dfirst-order}
\end{align}
where $\left\langle \varphi  \right|\hat A\left| \varphi  \right\rangle$ is the expectation value of ${\hat A}$ in the state $\varphi$, and ${d_{\rm{in}}}$ denotes the mean value of the input mode.

The variance of the final data can be obtained and is
\begin{align}
	{\sigma ^2} &= \langle \gamma |\hat D_{{t_j}}^{{\rm{SNU}}}\hat D_{{t_j}}^{{\rm{SNU}}}{\left| \gamma  \right\rangle _{{\xi _{\rm{A}}}}} - \langle \gamma |\hat D_{{t_j}}^{{\rm{SNU}}}\left| \gamma  \right\rangle _{{\xi _{\rm{A}}}}^2   \notag  \\
	& = \eta \langle \gamma |{{\hat X}_{{\xi _{\rm{A}}}}}{{\hat X}_{{\xi _{\rm{A}}}}}{\left| \gamma  \right\rangle _{{\xi _{\rm{A}}}}} + 2\sqrt {\eta \left( {1 - \eta } \right)} \langle \gamma |{{\hat X}_{{\xi _{\rm{A}}}}}{{\hat X}_{{\Psi _ \bot }}}{\left| \gamma  \right\rangle _{{\xi _{\rm{A}}}}}   \notag\\
	&+ \left( {1 - \eta } \right)\langle \gamma |{{\hat X}_{{\Psi _ \bot }}}{{\hat X}_{{\Psi _ \bot }}}{\left| \gamma  \right\rangle _{{\xi _{\rm{A}}}}}  \notag \\
	&= \eta V_{\rm{in}} + \left( {1 - \eta } \right) \cdot 1 ,   \label{Vsecond-order}
\end{align}
where $V_{\rm{in}}$ is the variance of the input mode.

From Eq.~(\ref{MeasurementResult_ATT}), we can see that the measurement results are equivalent to a mode-matching loss added before the receiver side, which is modeled by a beam splitter (BS). After the first-order and second-order moments of measured data are given, it is more intuitive to see that the transmittance of equivalent BS is $\eta$. In the above discussion, we assume that $\left| \gamma  \right\rangle$ is a TM coherent state to simplify the calculation of Eqs.~(\ref{dfirst-order}) and (\ref{Vsecond-order}). While one can further exam that, for an arbitrary input state, Eqs.~(\ref{dfirst-order}) and (\ref{Vsecond-order}) still hold. The above derivations also hold considering heterodyne detection.


\section{DATA AVAILABILITY}

The data that support the findings of this study are available from the corresponding author upon reasonable request.

\section{CODE AVAILABILITY}

The code used in this study is available from the corresponding author upon reasonable request.

\section{Acknowledgments}

This work was supported by the National Natural Science Foundation of China (Grants Nos. 62201012, 62001041, and 61531003), China Postdoctoral Science Foundation (Grant No. 2020TQ0016), the Fundamental Research Funds of BUPT (Grant No. 2022RC08), and the Fund of State Key Laboratory of Information Photonics and Optical Communications (Grant No. IPOC2022ZT09).

\section{AUTHOR CONTRIBUTIONS}
All authors contributed to the scientific discussions and the theoretical developments of the study. Z.C. and Z.L. carried out
the theoretical calculations, X.W. performed the simulation, Z.C. wrote the manuscript, and X.W., S.Y., Z.L., and H.G. provided revisions.

\section{COMPETING INTERESTS}
The authors declare no competing interests.

\section{ADDITIONAL INFORMATION}
Correspondence and requests for materials should be addressed to Ziyang Chen (chenziyang@pku.edu.cn), Zhengyu Li (lizhengyu2@huawei.com) and Hong Guo (hongguo@pku.edu.cn).

\end{document}